\documentclass[letterpaper, conference]{ieeeconf}  

\IEEEoverridecommandlockouts                              

\usepackage{graphicx} 
\usepackage{booktabs} 
\usepackage{url}
\usepackage{mathptmx} 
\usepackage{amsmath} 
\usepackage{amssymb}  
\usepackage{breqn}
\usepackage{enumerate}
\usepackage{dblfloatfix} 

\title{\LARGE \bf Three Approaches to the Automation of Laser System \\  Alignment and Their Resource Implications: A Case Study}

\author{David A. Robb$^{1}$, Donald Risbridger$^{2}$, Ben Mills$^{3}$, Ildar Rakhmatulin$^{2}$,
\\
Xianwen Kong$^{2}$, Mustafa Erden$^{2}$, M.J. Daniel Esser$^{2}$, Richard M. Carter$^{2}$ and Mike J. Chantler$^{1}$
\thanks{
For the purpose of open access, the author has applied a Creative Commons Attribution (CC
BY) license to any Author Accepted Manuscript version arising. This work was funded by the the UK EPSRC  projects EP/V051164/1, EP/T026197/1 and EP/W028786/1.}
\thanks{$^{1}$School of Mathematical and Computer Sciences, $^{2}$School of Engineering and Physical Sciences, Heriot-Watt University, UK, $^{3}$Optoelectronics Research Centre, University of Southampton, UK. (email: d.a.robb@hw.ac.uk).
}}

\begin{document}

\maketitle

\thispagestyle{empty}
\pagestyle{empty}

\begin{abstract}

The alignment of optical systems is a critical step in their manufacture. Alignment normally requires considerable knowledge and expertise of skilled operators. The automation of such processes has several potential advantages, but requires additional resource and upfront costs. Through a case study of a simple two mirror system we identify and examine three different automation approaches. They are: artificial neural networks; practice-led, which mimics manual alignment practices; and design-led, modelling from first principles. We find that these approaches make use of three different types of knowledge 1) basic system knowledge (of controls, measurements and goals); 2) behavioural skills and expertise, and 3) fundamental system design knowledge. We demonstrate that the different automation approaches vary significantly in human resources, and measurement sampling budgets. This will have implications for practitioners and management considering the automation of such tasks.

\end{abstract}

\section{INTRODUCTION}

Optical systems such as telescopes, interferometers, spectrometers and lasers often require skilled alignment during assembly in manufacture and installation \cite{hecht2012optics, rakhmatulin2023addressing, hinrichs2020neural}. Incorrectly aligned systems perform sub-optimally, if at all \cite{hecht2012optics}. Industry and researchers are seeking to automate \cite{hinrichs2020neural}, speed up or create automated tools to help with \cite{kim2007merit,oteo2013new}, alignment processes. However, developing the automation of such systems requires additional time and resources, making it expensive to set up. The automation of any process is only worthwhile if it results in a more cost effective system. The automation approach critically impacts on a) the human resource required and b) the sampling budget  needed to establish the automation. This latter aspect can have a considerable impact on efficiency and hence cost (e.g. taking a single alignment measurement might take a few minutes, multiplied by the number of measurements required). We therefore, in this paper, present a case study of different automation approaches, using a simple but representative two-mirror alignment process (described in Section \ref{subsec: The alignment automation problem}). The three automation approaches are: 
\begin{itemize}
    \item \textbf{Neural network (Approach 1)}, training an artificial neural network (ANN) model and then predicting the mirror adjustment required for alignment.
    \item \textbf{Practice-led (Approach 2)}, which in essence adopts the alignment strategy applied by skilled human
    operators.
    \item \textbf{Design-led (Approach 3)}, exploiting a mathematical model of the system in calculating the mirror adjustment required for alignment.
\end{itemize}
For each approach we discuss (i) the human resource required to design and test the system and hence the likely implementation effort and (ii) the typical sampling budget required (the number and cost of instrument readings). The trade-offs between i) and ii) will have a bearing on the overall cost of any subsequent production run, thus informing any cost-benefit analysis \cite{SHEHAB2001341}.

To the best of our knowledge the automation of laser system alignment has not been examined in this way before.

\textbf{The contributions of this work are: }
\begin{enumerate}
    \item Demonstration of the automation approaches:
    \begin{enumerate}
        \item A neural network alignment solution for the system.
        \item An observation study of alignment practitioners which exposes and characterises actual alignment practice on the example system, informing one of the automation approaches.
        \item A linear regression solution for the example system.
    \end{enumerate}
    \item Discussion contrasting the different elements of the implementation effort across the three automation approaches, including the different human resources (roles and skills) and knowledge categories which require to be traded off against the sampling budget (the number and cost of each alignment measurement reading which a given approach requires).
\end{enumerate}

The rest of this paper is organised as follows: In the next section we briefly describe prior work relating to the automation approaches for alignment of optical systems. In Section \ref{sec: Definition of automation problem } we portray the example system, define the automation problem and provide regression modelling terminology important for understanding this work. We then set out three approaches for automation: first, in Section \ref{sec: A New Neural Network Solution}, a new neural network solution; secondly, in Section \ref{sec: practice led Automated Solution} an automated alignment solution emulating manual alignment practice; and thirdly, in  Section \ref{sec: A design knowledge led approach}, a linear regression approach. In Section \ref{sec: Discussion}, we compare these different approaches and point out limitations. Finally, we summarise this case study, suggesting directions for future work.
\vspace{1mm}

\section{Background}
\label{sec: Background}
Various machine learning approaches have been applied to optical systems due to their multi-dimensional alignment spaces often making manual alignment challenging and time intensive \cite{tunnermann2019deep}, \cite{RAKHMATULIN2024107923}. These approaches seek to avoid the shortcomings of prior design-led solutions to these alignment problems/calibrations, including the inability to automatically account for even small system changes, their reliance on user expertise and their susceptibility to system errors, particularly when compared to regression methods \cite{7128690}.

Neural networks have been widely applied across this field due to their ability to solve nonlinear problems. They have been used for fast-axis collimation of laser diodes \cite{Khachikyan:21}, galvanometric laser scanning \cite{s18010197} and in the optimisation of deep ultraviolet mode-locked lasers \cite{article} to name but a few application areas.

Alternatively, an example of an automated alignment method supplementing known manual techniques is the Raspberry Pi auto-aligner system 
\cite{Mathew_2021}. This setup used the open source machine learning algorithm M-LOOP (specifically the Gaussian processes) with stepper motors to improve upon the manual alignment (done using the ``walking the beam'' method described in Section \ref{sec: practice led Automated Solution}) of a laser beam into a single-mode optical fiber and to perform continual optimisation, as opposed to aligning from start to finish. Another practice-led automation method is used in 
\cite{SalazarSerrano2018HowTA}, where stepper motors were used to generate a look up table for coupling light into a multi-mode optical fiber, from which the motor positions are then set to those corresponding to the maximum transmitted power.

Consequently, while there are many examples of neural network optical alignment solutions, 
fewer published works investigate practice-led or design-led automation solutions. 
We found no work comparing these three categories of automation in terms of their human resource, knowledge resource and sampling budget requirements.
\vspace{1mm}

\section{Definition of the automation problem and regression modelling terminology}
\label{sec: Definition of automation problem }

\begin{figure}
    \centering
      \includegraphics[scale=0.3]{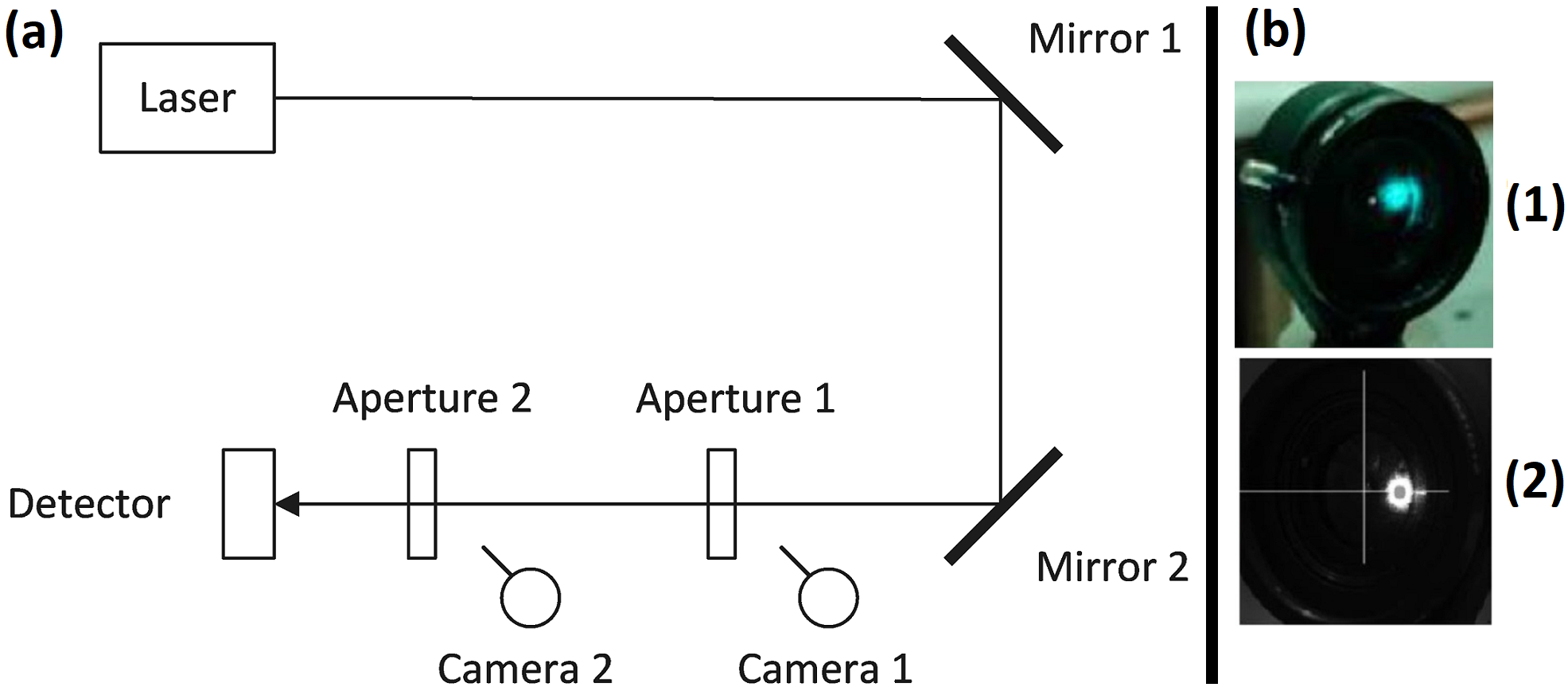}
      \caption{ (a) Schematic diagram of the system showing components including two computer vision cameras. (b) An aperture with incident laser beam (1) and the corresponding processed computer vision image allowing precise location of the beam in relation to the aperture centre (2). 
      }
      \label{fig:sys_inc_cameras}
\end{figure}

\subsection{The example system}
\label{subsec: The example system }
\begin{figure}
    \centering
      \includegraphics[scale=0.75]{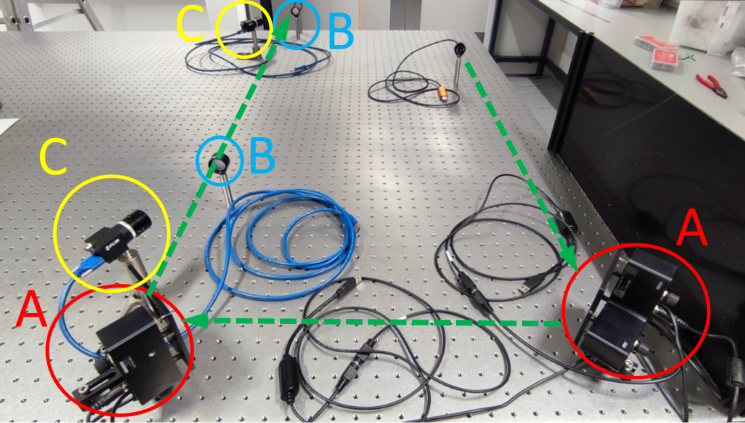}
      \caption{Implementation of the example system. The dashed line traces the laser beam path from the laser source to the collector/power meter. A) motorised mirror mounts, B) apertures and C) computer vision cameras.
      }
      \label{fig:motorisedSystem}
\end{figure}

We chose to base the case study on a simple but representative two-mirror system. (See Fig. \ref{fig:sys_inc_cameras}). 
It is common for optical systems to include mirrors so as to facilitate alignment of a laser beam's path \cite{heintzmann2013practical}. Thus, we included, in addition to a laser source, two mirrors in adjustable two-axis (pitch and yaw) mounts. To add a system constraint that would necessitate precise alignment, two apertures were included to define a specific path (or optical axis) of the beam on its way to a collector, which took the form of a power meter. This form of problem, steering a beam along an optical axis using two mirrors, is a particularly common alignment task \cite{heintzmann2013practical}. We include computer vision cameras as these allow measurement of the displacement of the beam from the centre of each aperture. 
The system with its motorised kinematic mirror mounts and computer vision cameras allowed both automation and computer controlled measurement data sampling. It was built using off-the shelf optical components
(Fig. \ref{fig:motorisedSystem}).

\subsection{The alignment automation problem}
\label{subsec: The alignment automation problem}

\vspace{1mm}

\begin{figure}
    \centering

      \includegraphics[scale=0.9]{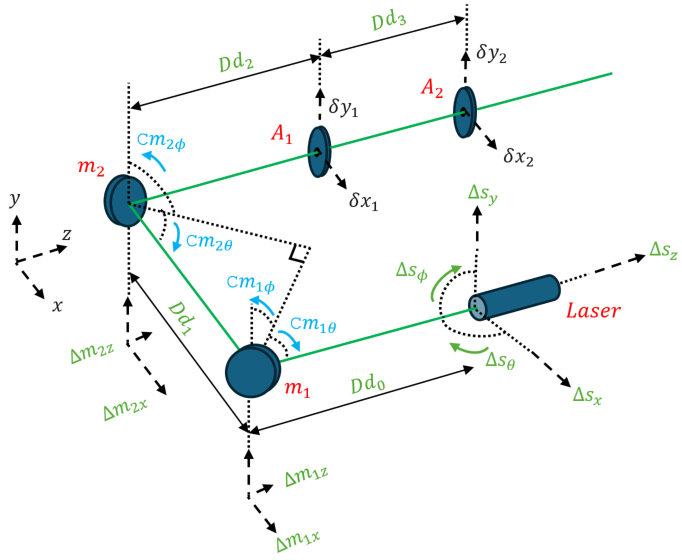}
    \caption{ Diagram modelling the example system in three dimensions. The laser beam emanates from a laser and encounters the components: first mirror $m_1$, then mirror $m_2$, next aperture $A_1$ and finally aperture $A_2$. The layout and notaion is inspired by \cite{Yuan:11}. 
    }
      \label{fig:3DSystemDiag}
\end{figure}
\vspace{1mm}

We define the system in three dimensions in Fig. \ref{fig:3DSystemDiag} and the automation problem below in terms of the components, measurements, the beam-blocking issue, controls and goal.
\vspace{1mm}

\textbf{Components}: The system has a laser, two mirrors and two apertures.

\textbf{Measurements}: On the apertures the displacement of the incident laser beam from the centres are measured by computer vision cameras. Each measurement can be described thus:  ($\delta$$x_1$, $\delta$$y_1$, $\delta$$x_2$, $\delta$$y_2$) where $\delta$$x_1$ is the displacement of the beam in the horizontal axis from the centre of Aperture 1 and so on. 

\textbf{Beam-blocking issue}: We assume that we always get x,y measurements, ($\delta$$x_1$, $\delta$$y_1$), from Aperture 1, but that we do not initially necessarily get measurements, ($\delta$$x_2$, $\delta$$y_2$), from Aperture 2, due to Aperture 1 sometimes blocking the beam\footnote{This is a common issue in optics alignment and it is standard practice to align components sequentially starting from the laser source and working downstream until all are aligned}.

\textbf{Controls}: Mechatronic controls for the mirrors are defined as ($Cm_1$$_\theta$, $Cm_1$$_\phi$, $Cm_2$$_\theta$, $Cm_2$$_\phi$) where $Cm_1$$_\theta$ represents the yaw\footnote{While often in optics practice the terms ``tip'' and ``tilt'' are used, those terms can be ambiguous, therefore here we use ``pitch'' and ``yaw'' to avoid ambiguity. These correspond to $\theta$ and $\phi$ respectively in Fig. \ref{fig:3DSystemDiag}.} angle and $Cm_1$$_\phi$ the pitch angle adjustment for Mirror 1 and so on. (See Fig \ref{fig:3DSystemDiag}).

\textbf{Goal}: The alignment objective is (0,0) on both apertures. This will equate to a measurement value, ($\delta$$x_1$, $\delta$$y_1$, $\delta$$x_2$, $\delta$$y_2$), of (0,0,0,0).
\vspace{1mm}

\subsection{Forward and reverse regression model terminology}
\label{subsec: Forward and reverse model terminology}
Subsequent sections describe the application of regression modelling. To aid in understanding we clarify here that a forward model is one which takes inputs as predictors (in this case mirror control adjustments) and models the outputs as outcomes (in this case measurements of the displacement of the beam from aperture centres) \cite{field2009discovering}. Conversely, a reverse model takes ``outputs'' (measurements) as predictors and models ``inputs'' (mirror controls) as the outcomes.  

\section{Neural Network (Approach 1)}
\label{sec: A New Neural Network Solution}

\begin{figure}
    \centering
     \framebox{
      \includegraphics[scale=0.18]
      {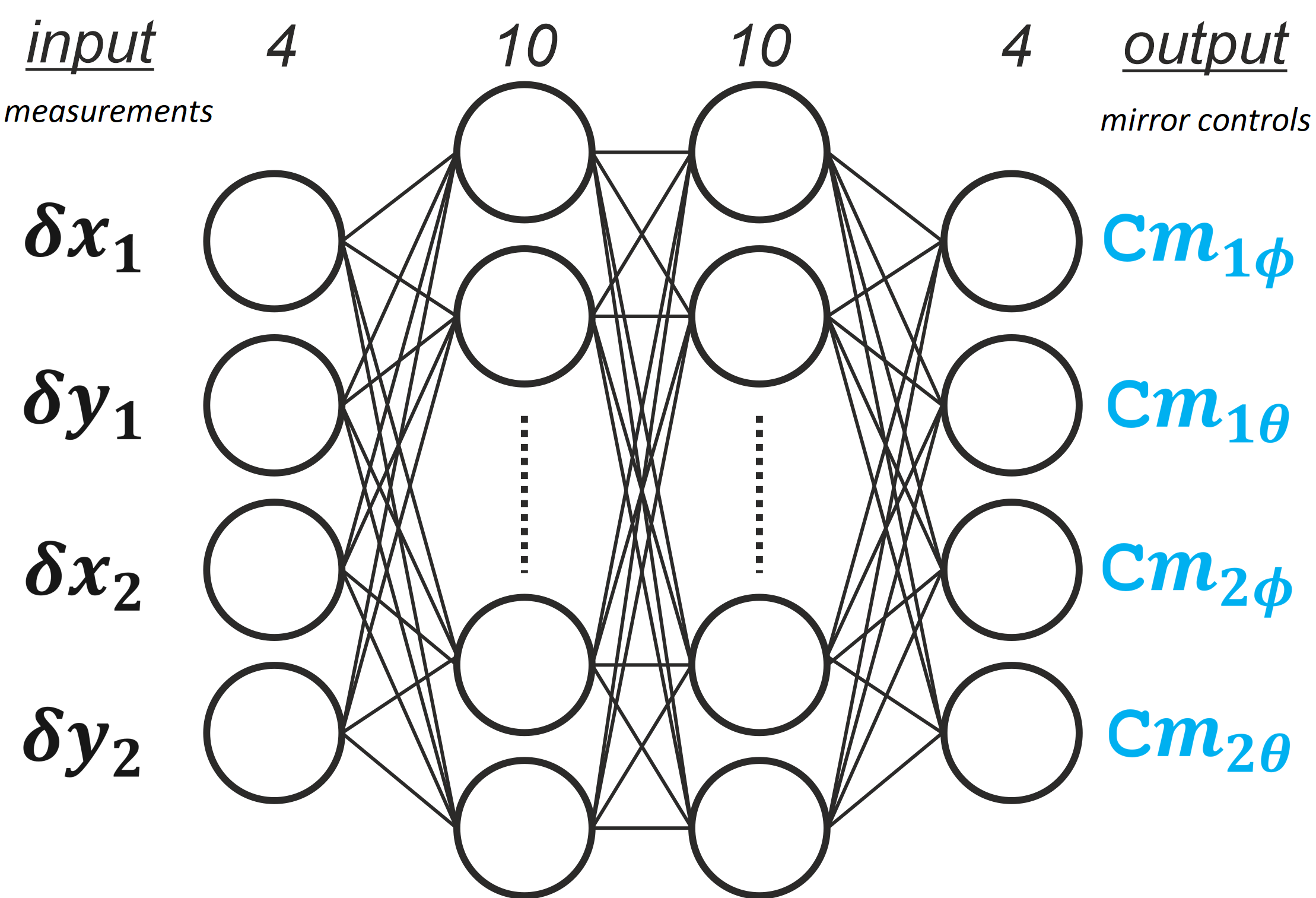}}
      \caption{Diagram of the ANN trained in the reverse model with the aperture 
      measurements as inputs and the mirror controls 
      as outputs. 
      (See Fig. \ref{fig:3DSystemDiag}).}
      \label{fig:ANNDiag}
\end{figure}

\begin{figure}
    \centering
      \includegraphics[scale=0.5]
      {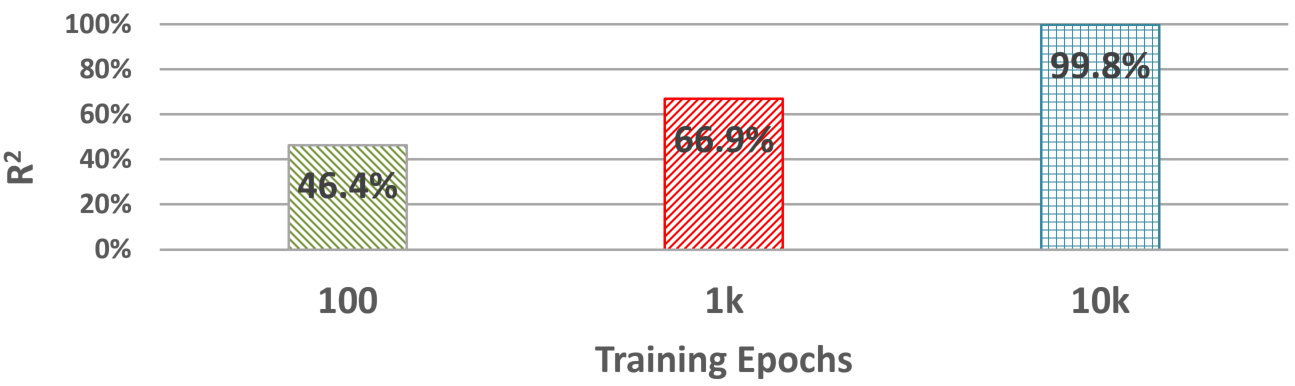}
      \caption{Chart of ANN training epochs v.s. mean $R$$^2$ model goodness of fit showing that training for 10000 epochs produced the highest goodness of fit to the training data (see text).}
      \label{fig:ChartRsquaredVsEpochs}
\end{figure}

\textbf{Overview}: We use an artificial neural network (ANN) to derive a reverse (possibly non-linear) regression model, 
 ($Cm_1$$_\theta$, $Cm_1$$_\phi$, $Cm_2$$_\theta$, $Cm_2$$_\phi$) = f$_a$$_n$$_n$
 ($\delta$$x_1$, $\delta$$y_1$, $\delta$$x_2$, $\delta$$y_2$), i.e. a model that, given specified measurements ($\delta$$x$,$\delta$$y$), can estimate the required mirror control adjustments ($Cm$). This is a simple ``brute force'' approach in which the alignment space is randomly sampled, ignoring the beam-blocking issue (and setting aside incomplete samples).
 \vspace{1mm}

\textbf{Required knowledge}: Samples of inputs (controls), outputs (measurements) and the goal. (See Section \ref{sec: Definition of automation problem }).
\vspace{1mm}

\textbf{Method}: 
Using the motorised version of the system we randomly sampled control settings ($Cm$) 
from the alignment space and collected corresponding measurements, ($\delta$$x$,$\delta$$y$). 
The sampling confirmed the assumption that we always get measurements,  ($\delta$$x_1$, $\delta$$y_1$), from Aperture 1 (Section \ref{sec: Definition of automation problem }). However, samples where measurements, ($\delta$$x_2$, $\delta$$y_2$), for Aperture 2 were not obtained (i.e. blocked by Aperture 1) were set aside. The sample consisted of 1000 measurements 375 being set aside for this reason. The remaining 625 measurements were used to train and test a reverse ANN model i.e. f$_a$$_n$$_n$
 ($\delta$$x_1$, $\delta$$y_1$, $\delta$$x_2$, $\delta$$y_2$) (Fig. \ref{fig:ANNDiag}) which, in practice could be used to obtain the solution, 
($Cm_1$$_\theta$, $Cm_1$$_\phi$, $Cm_2$$_\theta$, $Cm_2$$_\phi$) = f$_a$$_n$$_n$
 (0,0,0,0).  
\vspace{1mm}

Python and PyTorch were the tools used. The ANN architecture was a regression network with 4 inputs, 4 outputs and 2 hidden layers each with 10 neurons (Fig. \ref{fig:ANNDiag}). ReLU was used after each layer. 
The settings used were: Optimiser: ADAM; Loss: MSE, and Batch size for training: 10. Training was run for 10,000 epochs. Fig. \ref{fig:ChartRsquaredVsEpochs} explains this choice. 
Prior to training, the samples containing full measurements were shuffled (to create a random order). Then the first 90\% of the rows were used as training data and the last 10\% as testing data. The network was trained with the inputs being the measurements, ($\delta$$x_1$, $\delta$$y_1$, $\delta$$x_2$, $\delta$$y_2$),  and outputs being the controls, ($Cm_1$$_\theta$, $Cm_1$$_\phi$, $Cm_2$$_\theta$, $Cm_2$$_\phi$). 
Therefore, once trained, one could simply set the inputs to (0,0,0,0) and then the neural network would make a prediction for the mirror positions corresponding to correct alignment through the apertures. However, rather than assess the trained model with a single input point, we calculated the model's goodness of fit metric, $R$$^2$ which describes the percentage of variance in the data (in this case the training data samples) explained by the model \cite{field2009discovering}. The model predicts each of the four controls and thus has four $R$$^2$ goodness of fit values. The mean $R$$^2$ value across the four 
values was 99.8\% meaning the model explained 99.8\% of the variance in the training data.

\vspace{1mm}

This ANN approach requires personnel familiar with neural network training techniques
. Also, computer coding skills were required to program the sample collection. This approach required the largest number of samples of the three approaches (1000 samples, 375 being set aside as they did not include measurements on both apertures).

\section{Practice-led (Approach 2)}
\label{sec: practice led Automated Solution}
\textbf{Overview}: We observed skilled alignment practitioners and, found that the more efficient strategy used was that termed ``beam walking'' \cite{heintzmann2013practical} which was applied in a way that took account of the beam-blocking issue. 
We implemented an algorithm to apply that efficient strategy.
 
\textbf{Required knowledge}: In addition to knowledge of the controls and the measurements 
and the alignment goal, this also required knowledge of the ``beam walking'' strategy possessed by skilled alignment practitioners. 

\textbf{Method}:
This approach was developed by first setting up a manually operated version of the example system with which we could study how optics experts actually align the system. We wrote a standard operating procedure (SOP) which reflected this practice. Based on this we created an automated solution for aligning the system to mimic the SOP inspired by human operators. These steps are described in the subsections below:
\vspace{3mm}

\subsection{A manually operated version of the example system}
\begin{figure}
    \centering
      \includegraphics[scale=0.7]{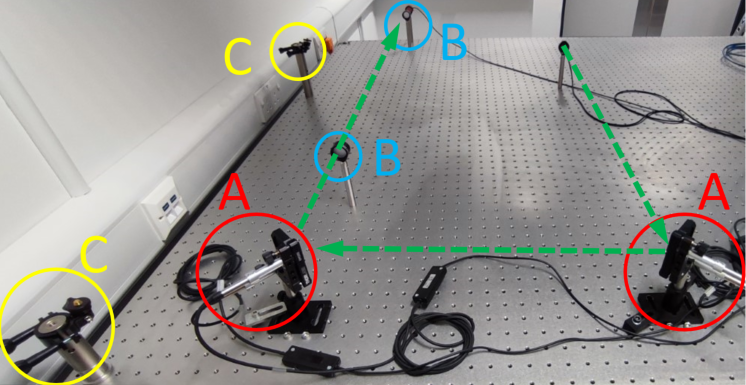}
      \caption{Manual version of the system. Labelled components: A) In contrast to the motorised system
      , the manual mirror mounts were customised with sensors (the silver tubes protruding from the rear of the mounts) for manual adjustment telemetry; however B) the apertures were identical to the motorised system; C) mounts for small observation video cameras. An additional video camera beyond the optical mounting table captured the participant's behaviour and ``think aloud'' speech audio (see Fig \ref{fig:expert_elan}).
      }
      \label{fig:ManualSystem}
\end{figure}

While the mirror mounts of the motorised system did allow manual adjustments, these were entirely non-intuitive controls that simultaneously adjusted both the angle and the velocity of angular adjustment of the mirrors. To properly study human alignment strategies a manually alignable version of the system was required. Therefore, we set up a duplicate system with manually adjustable kinematic mirror mounts in place of the motorised mounts.
Instead of computer vision cameras, optics experts use their eyesight to locate the beam on the apertures. We added mirror mount telemetry in the form of sensor probes to detect the amount of adjustment of yaw and pitch for each mirror mount as well as observation video cameras (see Fig. \ref{fig:ManualSystem}). This enabled the capture of human adjustments and alignment strategies. 
A power meter was included to facilitate manual alignment as human eyesight was not acute enough to accurately judge beam position on Aperture 2 when standing by the mirrors to make alignment adjustments (at 1.2m distance from Aperture 2). The power meter readings allowed confirmation that the beam was passing through Aperture 2. This was not needed to achieve alignment of the automated, motorised system using the algorithm that was eventually implemented. 
The key parameters, the distances between the components ($Dd$$_0$ to $Dd$$_3$ in Fig. \ref{fig:3DSystemDiag}) were identical in both systems.
\vspace{4mm}

\vspace{1mm}

\subsection{How optics experts align the system}
\label{subsec: How expert aligners align}

\begin{figure*}
    \centering
      \includegraphics[scale=0.73]{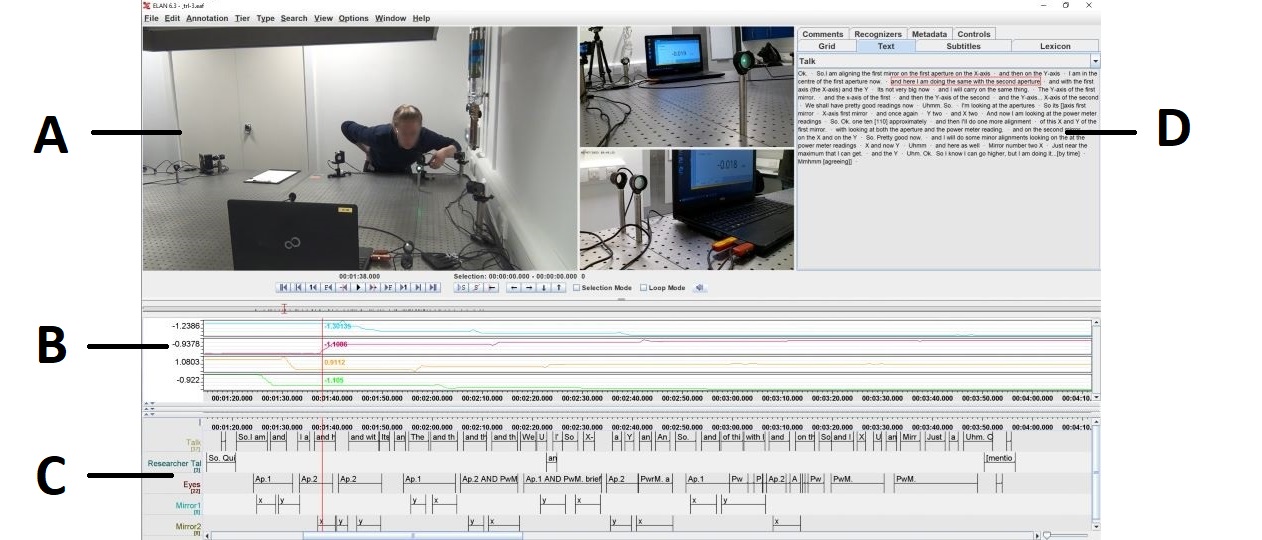}
      \caption{Illustrative screen of a laser alignment session being analysed showing the layout of the facilities in the ELAN synchronisation and annotation software. (Fig. \ref{fig:beamwalk} shows detail). A) media (video) panel with synchronised views (note the laptop used to display the output of the power meter for the participant to view); B) data trace panel with mirror telemetry; C) annotation tracks; D) text transcript panel.  Fig. \ref{fig:beamwalk} shows a close detail view of B and C. }
      \label{fig:expert_elan}
\end{figure*}

To investigate how optics practitioners actually align the example system, we recruited five optics experts from within our institution, but outside our project team, to observe how they actually carried out alignment. Their average age was 28; 3 identified as female, 2 as male. They reported their length of experience working with lasers as follows: one had over ten years, three had three to five years, and one had one to two years of experience. 

We developed an observation and interview protocol and 
piloted it on one of our team's own optics experts. This allowed the refinement and scoping of the alignment task and the number of sub-tasks with regard to duration so as to minimise participant fatigue. 

The recruitment procedures and protocols were approved by the Ethics Committee of the School of Mathematical and Computer Sciences, Heriot-Watt University, Edinburgh, UK.
\vspace{-1mm}

\textbf{The Alignment Sessions:} The protocol for the alignment sessions, interviews, and their analysis was as follows:
\begin{enumerate}
    \item Induction: Protocol and data recording were explained with opportunities for questions. Participant gave written informed consent.
    \item The system and the alignment task were explained and discussed; the task being to realign the system (after deliberate misalignment by the researcher) such that the maximum power would be achieved at the collector. 
    \item Two or three recorded alignment trials (time permitting), one of which would be time constrained 
    so as to expose possible short cut solutions to an alignment.

\end{enumerate}
Participants were encouraged to ``Think Aloud'' (i.e. to verbalise their thinking \cite{priede2011comparing, WOLCOTT2021181}) while doing their alignment tasks. In addition, they were allowed to ask questions of the researcher at any stage during the alignment session.

\textbf{Alignment session analysis:}
We used the ELAN annotation tool \cite{brugman2004annotating, aguera2011elan}, to annotate the actions and behaviours of the expert participants. ELAN allows annotation of observed actions in multimedia recordings \cite{sandgren2014coordination}. It also allows the synchronisation of multimedia recordings and data streams such as the telemetry readings from the manual system's modified kinematic mirror mounts (See Fig. \ref{fig:ManualSystem}). The synchronised data can then be annotated along the time dimension (see Fig. \ref{fig:expert_elan}). ELAN also enables the annotator to add text transcription of the audio recording in an annotation track (or 'tier'), and so include the participant's verbalised commentary from their ``thinking-aloud'' during the task.

During the analysis after each alignment session, episodes of interest 
were identified to be used to focus the subsequent follow-up interview such that strategies, behaviours, patterns of control adjustments, and participants' verbalised thoughts could be probed. E.g. participant 2 (P2) was probed on the concept of ``range'' having verbalised about this during 
the task when trying to establish how much they could adjust a particular control on a particular mirror.

\begin{figure}
    \centering
      \includegraphics[scale=0.55]{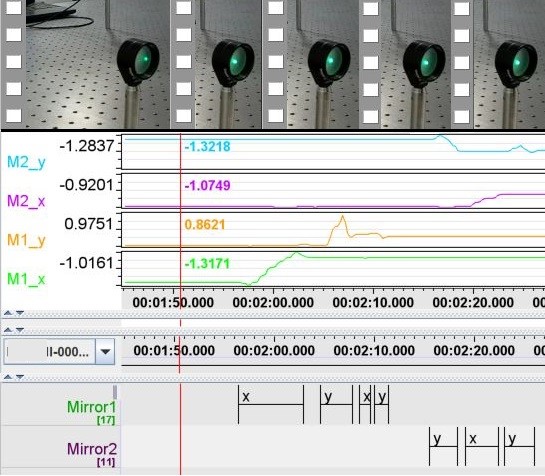}
      \caption{``Beam walk'' alignment practice. Shown by a montage of, a screen capture from the ELAN analysis of the first 30 seconds of a trial, along with (top) cropped frames from the observation video of Aperture 1. The ELAN screen shows the telemetry traces for the mirror adjustments e.g. Mirror1 x-axis (labelled M1\_x in ELAN), 
      and annotations for Mirror 1 and Mirror 2 yaw and pitch controls as they are adjusted (labelled x and y). It shows how a participant ``walks'' the beam sequentially, horizontally and vertically across the aperture surface, targeting the aperture centre, by adjusting 
      M1\_x, M1\_y, (followed by small adjustments of M1\_x and M1\_y again), then M2\_y, M2\_x and lastly a small adjustment of M2\_y.}
      \label{fig:beamwalk}
\end{figure}
\vspace{-1mm}

\begin{figure*}[b]
    \centering
      \includegraphics[scale=0.875]{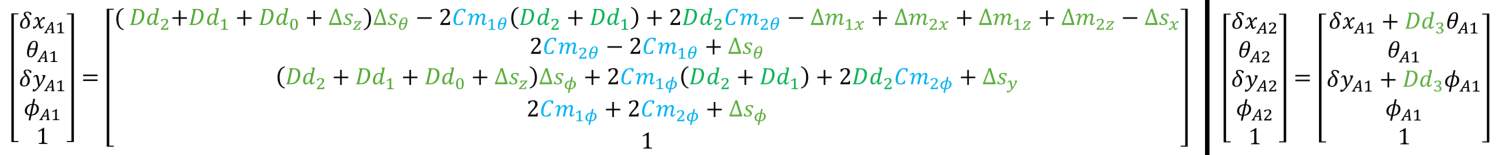}
      \caption{The two equations, together modelling the example system. This accompanies the diagram in Fig \ref{fig:3DSystemDiag}.  $C$ represents the mirror controls, adjusted during alignment (e.g. $Cm$$_1$$_\theta$ is the pitch adjustment for Mirror 1). $D$ represents the defined parameters, which in this case are the distances between the optics, with $\delta$ representing the displacement of the beam on the apertures in the x,y plane in relation to the optimum position, perfect alignment is therefore represented by the vector [0,0,0,0,1] on the apertures. $\Delta$ represents the “errors” i.e. the parameters that change between different instances of the system (were others to be constructed). Fig. \ref{fig:3DSystemDiag} diagram (and notation) uses a constant x,y,z coordinate while the angles of rotation reflect the local frame of the component in reference to the designed beam path as it would be adjusted or measured in the application. }
      \label{fig:EquationFor3DDiag}
\end{figure*}
\vspace{-1mm}

\textbf{Follow-up interview session:}
The follow-up interviews used a semi-structured interview methodology \cite{Corbin08, Strauss87}. Participants were probed about the details of and reasons for their actions
. This probing was structured around a small number (3 or 4) of episodes of interest 
(described above).

\textbf{Results from expert alignment observations:} 

A particular methodical approach to the alignment was applied by all the participants for at least some portion of their alignment session. The technique is a recognised alignment technique for systems with two mirrors often termed, a ``beam walk'' which is a method of adjustment of the path of the beam 
\cite{heintzmann2013practical}. 
A segment from a participant's alignment record illustrates how the approach manifested itself in the telemetry and observation records  (see Fig. \ref{fig:beamwalk}). To successfully achieve alignment participants would deal with the beam-blocking issue (see Section \ref{sec: Definition of automation problem }) by sequentially focusing first on aligning on the centre of Aperture 1, then Aperture 2 until their adjustments resulted in beam-blocking at which point they would switch focus back to Aperture 1, and so on until alignment was achieved.

Some participants also used a 
less rigid approach 
resorting to relying on power meter readings early rather than using incident beam positions on the apertures, particularly when the initial misalignment was gross. The rigid sequential iterative approach resulted in swifter solutions (the fastest solution during a trial being achieved in 1 minute and 20 seconds). The less rigid approach usually took longer to reach a solution (e.g. in one trial it took 13 minutes to achieve a solution, but starting from a gross misalignment).

\subsection{A Standard Operating Procedure (SOP)}
\label{subsec:A Standard Operating Procedure}

We created a Standard Operating Procedure (or SOP) for the manual version of the system reflecting the more efficient of the observed approaches, i.e. the  iterative approach applying the ``beam walk'' technique. The SOP is set out in the steps below and refers to components shown in Fig. \ref{fig:sys_inc_cameras}.
\begin{enumerate}
    \item 
    Using the controls on Mirror 1, adjust the yaw and pitch of the beam such that the beam is aligned to the centre of Aperture 1.  
    \item 
    Similarly, using the controls on Mirror 2 adjust the beam such that it is aligned to the centre of Aperture 2, or it becomes blocked by Aperture 1.
    \item Repeat steps 1)  and 2) until no further adjustments are required. The system is aligned when the beam passes with maximum transmitted power through both apertures (as measured by the power meter)\footnote{The number of iterations is related to the length of $Dd$$_2$ (distance between 2nd mirror and 1st aperture) if this is zero no iteration would be needed. See Fig. \ref{fig:3DSystemDiag}.}.  
\end{enumerate}

\vspace{1mm}

\subsection{The practice-led automated solution}
The automated alignment solution that we developed is published in \cite{rakhmatulin2023addressing} and is inspired by the SOP which reflects the more time efficient of the strategies used by the 
participant group described above.

Similar to the SOP, it commences by measuring the distance that the beam is offset from the centre of Aperture 1, ($\delta$$x_1$, $\delta$$y_1$), see Fig. \ref{fig:sys_inc_cameras}, and adjusts the yaw and pitch of Mirror 1 until ($\delta$$x_1$, $\delta$$y_1$) is within a threshold distance of (0,0). It then measures the beam offset from the centre of Aperture 2 ($\delta$$x_2$, $\delta$$y_2$) and adjusts the yaw and pitch of Mirror 2, ($Cm_1$$_\theta$, $Cm_1$$_\phi$), until  ($\delta$$x_2$, $\delta$$y_2$) is within a threshold distance of (0,0) or until the beam is blocked from Mirror 2 by Aperture 1 (thus taking account of the beam-blocking issue). These steps are repeated until the beam offsets on Mirrors 1 and 2 are within a set threshold which represents a good alignment solution, at which point the automated adjustment ends. 

A typical run of this system involved a total of 190 computer vision camera readings before the algorithm terminated with a good alignment solution.

In addition to the personnel possessing practical knowledge of how to align the system manually, skills in knowledge and behaviour capture may be required. Programming skills were also needed to mimic the human strategy.

\section{Design-led (Approach 3)}
\label{sec: A design knowledge led approach}

\begin{table*}
  \centering
  \begin{tabular}{*{7}{c}}
    \toprule
    \textbf{Automated} &  \textbf{ Domain knowledge}  & \textbf{Additional automation} &   \textbf{No. of samples} & \textbf{Approach to}\\
    \textbf{Approach }&   \textbf{and expertise required} &  \textbf{knowledge required}&   \textbf{required}& \textbf{beam-blocking issue}\\
    \midrule
    Neural network & Inputs (controls), outputs   & Knowledge of generic regression  &     10\textsuperscript{3} & Brute force, discard\\
       &   (measurements) and goals &  and optimisation approaches&  &  beam-blocked samples\\
    \hline
    Practice-led &   Learned behaviour (skill), &  Knowledge of &    10\textsuperscript{2} & Informed by\\
      & production knowledge and expertise*&  search strategies  & &expert practice \\
    \hline
Design-led  &  Input-output mathematical    & Analytical and 
  &    10\textsuperscript{1}& Two-step process derived\\
       &  model*  &optimisation skills& &  from mathematical model\\

    \bottomrule
  \end{tabular}
    \caption{Human resource (knowledge), sampling required, and how the beam-blocking issue is addressed for each approach to automating a laser system. (*denotes: in addition to knowledge of the inputs, outputs and goals).}
  \label{tab:3 levels of knowledge}
\end{table*}

\textbf{Overview}: Consulting an expert in optics we obtained a full forward model. 
See Fig. \ref{fig:EquationFor3DDiag}, which, in conjunction with Fig. \ref{fig:3DSystemDiag}, models the 
system. The model assumes the ``small angle approximation'' which allows trigonometric ratios to be simplified when the angle is small (less than 15 degrees), true for our purposes as our motorised mirror mounts are only capable of $\pm$ 5.27 degrees of adjustment per axis 
\cite{
hecht2012optics}.
We examined the model to determine the linearity of the control variables. This showed that a multi-linear regression to estimate the reverse model is feasible \cite{montgomery2021introduction}. However, this would need full measurement readings. Thus, to address the beam-blocking issue, we adopted a two-step process that uses a regression of a partial model to then guarantee that full measurements for both Aperture 1 and Aperture 2 are obtained.

\textbf{Required knowledge}: The mathematical model, the small angle approximation and its applicability to our application; mathematical knowledge sufficient to recognise that the equation modelling the system is linear in nature; knowledge of multiple linear regression modelling; a small number of samples from the alignment space.

\textbf{Method}: 
The equations in Fig. \ref{fig:EquationFor3DDiag} were derived via ray transfer matrices \cite{hecht2012optics}. The matrices for the mirrors are non-standard and were determined by reference to \cite{Yuan:11}. The motorised version of the system was used for the sampling of control settings and measurements. The sampling was programmed in Python. The linear regression modelling was carried out offline in Microsoft Excel. The alignment goal (stated in Section \ref{sec: Definition of automation problem }) of (0,0) on both apertures equates to a measurement value, ($\delta$$x_1$, $\delta$$y_1$, $\delta$$x_2$, $\delta$$y_2$), of (0,0,0,0). If we were to sample randomly in the alignment space some of the sample mirror adjustments would result in no measurement at Aperture 2 (see ``beam-blocking issue'' in Section \ref{sec: Definition of automation problem }). To avoid this aspect leading to wasteful sampling we adopted a two-step approach: Step 1) Derive the linear relationship between Mirror 1 and Mirror 2 to achieve ($\delta$$x_1$, $\delta$$y_1$) = (0,0); Step 2) Use the Step 1 relationship to obtain data samples, 100\% of which are guaranteed to produce measurements on both apertures. The measurements from Step 2 are then used to derive the reverse linear model:   ($Cm_1$$_\theta$, $Cm_1$$_\phi$, $Cm_2$$_\theta$, $Cm_2$$_\phi$) = f$_l$$_i$$_n$
 ($\delta$$x_1$, $\delta$$y_1$, $\delta$$x_2$, $\delta$$y_2$) \cite{montgomery2021introduction}.

To achieve Step 1 we did the following:
\begin{enumerate}[i]
    \item We opted to use 30 randomly positioned samples plus four registration samples in which only one of the four adjustments was changed to allow validation. This gave measurements and control settings from 34 samples. 
    \item For simplicity, using the sampled data, we estimated the forward model (using linear regression) for Aperture 1, i.e. $B$ $=$ $am_1$ $+$ $bm_2$ + $c$ 
    \item We derived the linear relationship between Mirror 1 and Mirror 2 given $B$ $=$ 0 thus: $m_1$ $=$ $dm_2$ $+$ $e$
\end{enumerate}

For Step 2 we did the following:
\begin{enumerate}[i]
    \item Taking the same randomly chosen ($Cm_1$$_\theta$, $Cm_1$$_\phi$) control settings for Mirror 1 we applied the linear relationship from Step 1 iii) generating  ($Cm_2$$_\theta$, $Cm_2$$_\phi$) for Mirror 2 such that the beam would pass through Aperture 1. 
    \item Using this set of 34 control settings ($Cm_1$$_\theta$, $Cm_1$$_\phi$, $Cm_2$$_\theta$, $Cm_2$$_\phi$) we took 34 measurements,  ($\delta$$x_1$, $\delta$$y_1$, $\delta$$x_2$, $\delta$$y_2$) all of which registered measurements on Aperture 2. In theory fewer (perhaps 8) would have been needed but we used more to compensate for possible system noise.
    \item Using this set of 34 complete measurements we derived the reverse linear model: ($Cm_1$$_\theta$, $Cm_1$$_\phi$, $Cm_2$$_\theta$, $Cm_2$$_\phi$) = f$_l$$_i$$_n$
 ($\delta$$x_1$, $\delta$$y_1$, $\delta$$x_2$, $\delta$$y_2$).
    \item In a similar way to Approach 1, rather than assessing the model on a single point, (0,0,0,0), we calculated the model's goodness of fit metric, $R$$^2$. As in Approach 1, the model predicts each of the controls and thus has four $R$$^2$ goodness of fit values. The mean $R$$^2$ value across the four 
    goodness of fit values was 97.2\% meaning the model explained 97.2\% of the variance in the data samples.
\end{enumerate}

In this section we have described how a design knowledge inspired automated alignment solution was created based on using linear regression modelling. A comparatively small number of samples was used (34 each for Steps 1 and 2). However, expert system design knowledge was needed to produce a mathematical model of the system, including knowledge of the small angle approximation. Mathematical knowledge sufficient to recognise the linearity of the model was also required. Finally, knowledge of multiple linear regression modelling was needed to choose an appropriate sampling strategy and create the regression model which could output the alignment solution.

\section{Discussion and Limitations}
\label{sec: Discussion}

\begin{figure}
    \centering
      \includegraphics[scale=0.55]{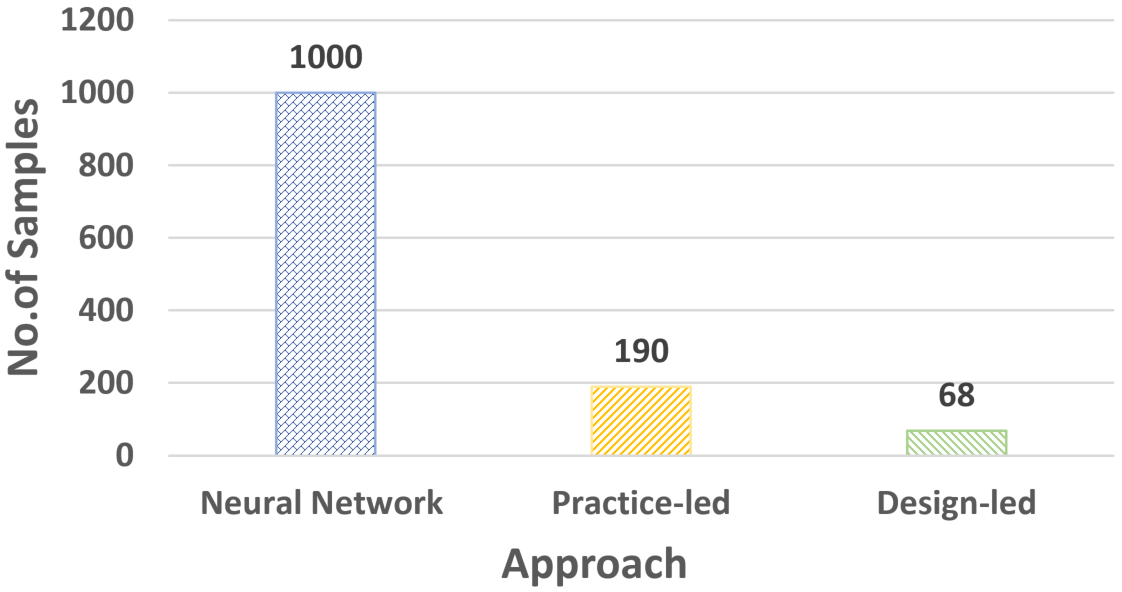}
      \caption{Chart showing how the sampling required varies for each approach.}
      \label{fig:NoOfSamples}
\end{figure}

In this section we contrast the three alignment automation approaches and discuss the implications of our observations about the resource requirements and sampling budgets associated with each for the automation of alignment in the manufacture of optical systems in general. We organise the main points in Table \ref{tab:3 levels of knowledge}. 

\textbf{The neural network approach} initially required the simplest knowledge as its basis, i.e. simply the control inputs (mirror adjustments), the outputs or measurements (taken using computer vision at the apertures), and the alignment goal itself, i.e. the beam reaching the centre of both apertures, (0,0,0,0). However, to implement this automation approach requires personnel expert in training neural networks and people with knowledge of generic optimisation approaches who can discriminate between optimisations that can and cannot be solved by a neural network. The overall implementation effort is potentially quite low as no detailed programming of the solution is required, simply the application of established neural network training techniques. However, the number of samples required is high (see Fig. \ref{fig:NoOfSamples}) and if a particular system's alignment measurements were time consuming and costly to obtain then this would be a clear disadvantage of the ANN approach. 

\textbf{The practice-led approach} requires a greater level of system knowledge at the outset in the form of the alignment strategies used by expert alignment practitioners. This type of information, of course, not only requires access to those who possess the knowledge but also to people who are able to  gather the expert knowledge from observing or interviewing practitioners. Thus for the practice-led approach the initial knowledge bar is higher than for the neural network approach. The programming skills required for this approach may also need to be wider than for the neural network approach to be able grasp a broad range of problems and their specific contexts, whilst also needing to be familiar with generic optimisation approaches. The more bespoke nature of each practice-led solution will mean that the implementation effort will be greater than for the more generic neural network approach. In favour of this approach though is the lower sampling budget needed to achieve alignment (Fig. \ref{fig:NoOfSamples}).

\textbf{The design-led approach} has the highest initial knowledge requirement of the three approaches. Personnel possessing the in-depth knowledge of the fundamental principles that drive the behaviour of any optical system, and can produce, understand and interrogate a mathematical model of that system are rare. The design and testing of the calculations which are required to inform the sampling, modelling and testing are time consuming and therefore expensive. Fewer samples are required to implement a design-led solution due to its targeted nature. This means that the sampling budget is low for this approach (34 for Steps 1 and 2, totalling 68 in this study, see Fig. \ref{fig:NoOfSamples}).

Importantly, the beam-blocking issue (a common aspect of many optical systems) which causes loss of data, is handled differently across the approaches. The neural network (Approach 1) simply discards the data in the incomplete samples. It is explicitly taken account of by practice-led Approach 2 informed by the experts' alignment strategies, while the design-led Approach 3 adopts a theoretically driven two-step process to handle the issue.

Bearing in mind the above and referring to Table \ref{tab:3 levels of knowledge}, we can see that the approach to alignment automation that should be adopted will depend on the factors of knowledge and skills availability, the availability of personnel and their roles 
and the sampling budget. For example: If measurements are cheap to take then it might be possible to attempt a generic neural network approach to see if a quick solution can be arrived at with that approach. This might be particularly attractive if modelling the system is expected to be difficult or expensive. If these aspects of knowledge availability and sampling budget are taken into account they can contribute as significant variables in any cost-benefit analysis for the design and manufacture of an optical system product.

As regards limitations, with this case study we have purposely chosen to limit the investigation scope to a simple example laser alignment task. However, it does present common alignment requirements that will be present in other systems. With regard to the three approaches: As approaches 1 and 3 are regression-based methods, unlike approach 2, they are not suited to problems which require local optimisation.

\section{Conclusions and Future work}
\label{sec: Conclusions and Future work}
We implemented a two-mirror laser system with motorised mirror mounts and computer vision cameras to use as a case study of laser alignment automation. Three automation approaches were explored, and each were found to have different knowledge requirements and hence human resource implications. Knowledge of control inputs,  output measurements, and the alignment goal informed a neural network. We studied how skilled alignment practitioners aligned a manual version of the system, noted the most efficient of the strategies and this knowledge informed a practice-led bespoke solution approach. Lastly, a design-led approach created a model and derived a regression solution. Additionally, the beam-blocking issue, common in optical system alignment (where upstream components must be aligned first to allow subsequent alignment of downstream components) was addressed differently in the three approaches. The differing types of initial system knowledge, personnel roles required for implementation, and sampling budgets across the three approaches, illustrate the trade-offs that should be made when embarking on automation of optical alignment processes. These would all affect the potential overall cost of implementing a system of automation of alignment tasks.

Future work in this direction will investigate examples of cost-benefit analyses and case studies on non-linear systems (such as telescope and$/$or resonator systems).

This paper, providing a novel comparison of approaches to the automation of laser alignment, has implications for the product development and manufacture of laser systems.










\bibliographystyle{IEEEtran} 
\bibliography{ourRefsGoHere}

\end{document}